\documentclass{webofc}

\usepackage[varg]{txfonts}   
\usepackage{hyperref}
\usepackage{url}
\hypersetup{colorlinks=true,citecolor=blue,urlcolor=blue,linkcolor=blue}

\usepackage{upgreek}
\usepackage{wasysym}

\newcommand{\be}{\begin{equation}}
\newcommand{\ee}{\end{equation}}
\newcommand{\ba}{\begin{align}}
\newcommand{\ea}{\end{align}}
\newcommand{\bea}{\begin{eqnarray}}
\newcommand{\eea}{\end{eqnarray}}

\newcommand{\F}{\mathcal{F}}
\newcommand{\FT}{\mathcal{F}^\mathrm{T}}
\newcommand{\Cl}{\mathrm{Cl}_2}
\newcommand{\Dl}{\Delta^{(1)}}
\newcommand{\Q}{\mathcal{Q}}

\newcommand{\Pih}{\hat\Pi}
\newcommand{\Pib}{\bar\Pi}

\newcommand{\amuSM}{\ensuremath{a_\upmu^\text{SM}}}

\newcommand{\amuexp}{\ensuremath{a_\upmu^\text{exp}}}

\newcommand{\amuexpresult}{116\,592\,071.5(14.5)}

\newcommand{\amuHVPtotalresult}{7\,045(61)}

\newcommand{\amuHLbLtotalresult}{115.5(9.9)}

\newcommand{\amuSMresult}{116\,592\,033(62)}
\newcommand{\amudiffresult}{38(63)}

\newcommand{\WPold}{Aoyama:2020ynm}
\newcommand{\WPnew}{Aliberti:2025beg}
\newcommand{\HSZ}{Hoferichter:2024bae}
\newcommand{\LMR}{Leutgeb:2022lqw}
\newcommand{\Tref}{Cappiello:2025fyf,Mager:2025pvz}

\begin{document}
\title{Hadronic light-by-light scattering in AdS/QCD and the
muon $g-2$: tensor meson contributions}
%
%

\author{        
        \firstname{Jonas} \lastname{Mager}\inst{1}\fnsep\thanks{\email{jonas.mager@tuwien.ac.at}}
\and
\firstname{Luigi} \lastname{Cappiello}\inst{2}\fnsep\thanks{\email{luigi.cappiello@unina.it}}
\and
\firstname{Josef} \lastname{Leutgeb}\inst{1}\fnsep\thanks{\email{josef.leutgeb@tuwien.ac.at}} 
\and
\firstname{Anton} \lastname{Rebhan}\inst{1}\fnsep\thanks{\email{anton.rebhan@tuwien.ac.at}}
        }

\institute{Institut f\"ur Theoretische Physik, Technische Universit\"at Wien,
        Wiedner Hauptstra\ss e 8-10,\\ 
        A-1040 Vienna, Austria \label{addr1}
        \and
        Dipartimento di Fisica ``E. Pancini", Universit\`a di Napoli ``Federico II", and
INFN-Sezione di Napoli, Via Cintia, I-80126 Napoli, Italy}


\abstract{%
We briefly review the predictions of holographic QCD 
in hard-wall AdS/QCD models for
the contribution of pseudoscalars and axial-vector mesons 
to the hadronic light-by-light amplitude and 
consequently to the muon anomalous magnetic moment,
which allows one to satisfy the Melnikov-Vainshtein constraint
in a purely hadronic model.
Moreover, we discuss the role of tensor mesons in satisfying
the remaining short-distance constraints, and provide details
for a minimal model of tensor mesons that turns out to
agree remarkably well with available data from BELLE as well as
preliminary low-$Q^2$ data from BESIII. 
In an appendix, we also provide analytical results
for the short-distance limits in hQCD and compare 
with the leading-order OPE result as
given by the massless quark loop.
}
\maketitle
\fancyhead[RO,LE]{\thepage}

\section{Introduction}

In June 2025, the Muon $g-2$ Experiment at the Fermi National
Accelerator Laboratory has released its final result
for $a_\upmu=(g-2)_\upmu/2$
with a precision of 127 ppb \cite{Muong-2:2025xyk}, which
provides the opportunity to test the Standard Model (SM) of particle
physics at an unprecedented level.
The most recent theoretical prediction from the
Muon $g-2$ Theory Initiative from May 2025 \cite{\WPnew} 
agrees within errors,
$\Delta a_\upmu\equiv\amuexp - \amuSM =\amudiffresult\times 10^{-11}$,
after a replacement of a data-driven determination of the
hadronic vacuum polarization (HVP) contribution by its
lattice prediction. Since the first White Paper (WP) of 2020
\cite{\WPold}, new data from $e^+ e^-$ colliders have
overthrown the previous data-driven determination, which
for the WP2025 prediction \cite{\WPnew} 
was replaced by a new lattice average, with an error budget
that is 50\% larger than the one assumed previously (and apparently
incorrectly). 

In order that the full potential as a test of Beyond-Standard-Model (BSM) physics
of the experimental measurement can be exploited, further progress
in the determination of hadronic contributions is necessary, and
is indeed the aim of ongoing research. 
Once the error has been
reduced significantly by further progress in lattice QCD and by
new experimental analyses in HVP,\footnote{A first notable reduction
of the uncertainty of the leading-order HVP contribution to $34\times 10^{-11}$ was obtained recently
in a hybrid approach which combines uncontested experimental data
with lattice QCD results \cite{Boccaletti:2024guq}.}
further refinements in the
already much improved determination of the hadronic light-by-light (HLbL)
contribution will also be called for.
Currently, there is a slight
tension between data-driven and lattice determinations of the latter, which
led to a scale factor of 1.5 in their combined error estimate,
indicating the need for further work, despite the fact that since the
first WP the uncertainty has already been reduced by nearly a factor of 2.

\begin{table}[t]
	\caption{Comparison of experimental and theoretical results
    for $a_\upmu\times 10^{11}$, and the hadronic contributions according to White Paper 2025 \cite{Aliberti:2025beg} and the previous numbers from 2020~\cite{Aoyama:2020ynm}} 
\small
\centering
	\begin{tabular}{lrr}
	\hline
	     & WP25\phantom{mmmnn} & WP20\phantom{mnn}\\ 
       \hline
Experimental world average & \amuexpresult & 116\,592\,089(63)\\
Standard Model prediction & \amuSMresult\phantom{nnn} & 116\,591\,810(43)\\
HVP & $\amuHVPtotalresult\phantom{nnn}$ & $6845(40)$ \\
HLbL 
&   $\amuHLbLtotalresult\phantom{n}$ & $92(18)$\\ 
        \hline
        \renewcommand{\arraystretch}{1.0}
	\end{tabular}
\label{tab:summary_comparison}
\end{table}

In the last years, holographic QCD (hQCD) has turned out to provide
a useful check of data-driven calculations of the HLbL contribution.
In particular, the best-guess hQCD predictions of Ref.~\cite{\LMR} (LMR22)
of axial-vector contributions and
related effects from short-distance constraints (SDCs) have turned
out to be confirmed remarkably closely by their first completely
dispersive treatment \cite{Ludtke:2024ase,\HSZ}. However,
it has become clear that the contribution of tensor mesons, which in fact play a
dominant role in photon-photon collisions, has been underestimated
significantly in the past. A new estimate based on the previously
employed simple quark model ansatz \cite{Schuler:1997yw},
but with kinematical singularities removed \cite{\HSZ}, predicted
a negative contribution, larger in size than found before \cite{Danilkin:2016hnh}.
On the other hand, using a (minimal) model of tensor-meson contributions in hQCD
we have found \cite{\Tref} even larger, but positive contributions, which in fact would
remove the tension of data-driven and lattice results for the total
HLbL contribution.\footnote{See ref.~\cite{Estrada:2025bty} for first, still inconclusive attempts to model tensor mesons within resonance chiral theory,
which are, however, consistent with the hQCD results.}

In this paper, we shall very briefly review the 
AdS/QCD hard-wall (HW) models and how they satisfy
SDCs, in particular the notorious Melnikov-Vainshtein (MV) \cite{Melnikov:2003xd}
constraint,
and how to a large extent the additional SDCs can be satisfied
by including minimally coupled tensor mesons.
In an appendix, we also provide analytical details of a comparison
with the full leading-order OPE results \cite{Melnikov:2003xd,Colangelo:2019lpu,Colangelo:2019uex,Bijnens:2020xnl,Bijnens:2021jqo}
given by the massless quark loop,
including a discussion of the leading terms in the corner
dynamics studied in \cite{Bijnens:2024jgh}.

\section{Hard-wall $N_f=2+1$ AdS/QCD model with $U(1)_A$ anomaly}

The hard-wall AdS/QCD model introduced in \cite{Erlich:2005qh} is formulated in pure 5-dimensional
anti-de Sitter space with line element
\be\label{ds2AdS}
ds^2=z^{-2}(\eta_{\mu\nu}dx^\mu dx^\nu - dz^2)
\ee
with conformal boundary at $z=0$
and a sharp cut-off in
the holographic coordinate at $z=z_0$.
Like the top-down holographic model of
Sakai and Sugimoto \cite{Sakai:2004cn} it
has nonabelian flavor gauge fields dual to left and right
chiral quark currents with a Chern-Simons action implementing
chiral anomalies, but additionally a bi-fundamental
scalar field $X$ dual to quark bilinear operators,
with action
\begin{align}
 \label{S5}
S =& -\frac{1}{4g_5^2} \int d^4x \int_0^{z_0} dz
\sqrt{g}\, g^{PR}g^{QS} \text{tr}\left(\mathcal{F}^\mathrm{L}_{PQ}\mathcal{F}^\mathrm{L}_{RS}
+\mathcal{F}^\mathrm{R}_{PQ}\mathcal{F}^\mathrm{R}_{RS}\right) 
+S_\mathrm{CS}[\mathcal{B}^\mathrm{L}]-S_\mathrm{CS}[\mathcal{B}^\mathrm{R}]
\nonumber\\
&+\int d^4x \int_0^{z_0} dz\,\sqrt{g}\;\text{tr}\left(
|DX|^2+3|X|^2 \right)-2k_T\int d^4x \int_0^{z_0} dz \sqrt{g}\,(R+2\Lambda)
\end{align}
where $P,Q,R,S=0,\dots,3,z$ and $\mathcal{F}_{MN}=\partial_M \mathcal{B}_N-\partial_N \mathcal{B}_M-i[\mathcal{B}_M,\mathcal{B}_N]$.
Setting $g_5^2=12\pi^2/N_c$ matches the asymptotic
behavior of the vector correlator to the
logarithm in the leading-order OPE result for QCD.

Quark masses can be incorporated through the 
boundary values $\lim_{z\to 0}X/z=\frac12\text{diag}(m_u,m_d,m_s)$,
and the chiral condensate appears in the asymptotics
of the normalizable modes of $X$.
This gives rise to masses for the pseudoscalars in
accordance with the Gell-Mann-Oakes-Renner relation.
Realistic masses for $\eta$ and $\eta'$ require, however,
an implementation of the U(1)$_A$ anomaly.
In \cite{\LMR} we have adapted the Katz-Schwartz model \cite{Katz:2007tf},
which introduces a complex scalar $Y$ as a dual
to the gluonic operator $\alpha_s(GG+iG\tilde  G)$
with a coupling  $\propto Y^{N_f} \det(X)$
and a background $Y_0$ that reflects the logarithmic
running of $\alpha_s$. Allowing for a nonvanishing
gluon condensate which was neglected in \cite{Katz:2007tf}
we found a remarkably good match of
the masses and two-photon-widths of $\eta$ and $\eta'$.
An alternative implementation of the U(1)$_A$ anomaly
in \cite{Leutgeb:2024rfs}
through a scalar-extended Chern-Simons action involving
so-called superconnections
turned out to be less satisfactory phenomenologically;
its deviations from the best-guess model in \cite{Leutgeb:2024rfs}
(LMR22v1) were taken as an indicator
of the theory error within the HW $N_f=2+1$ AdS/QCD models.

\begin{figure}
\centering
\includegraphics[width=0.5\textwidth,clip]{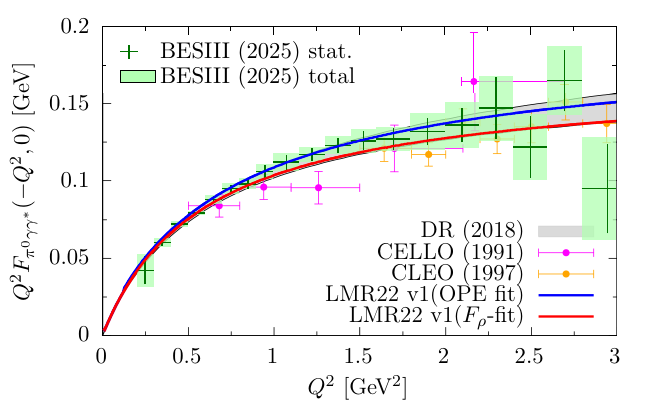}\includegraphics[width=0.5\textwidth,clip]{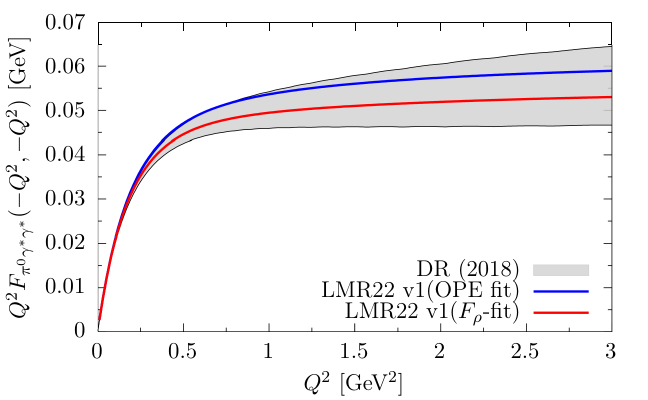}
\caption{Results for the singly and doubly virtual
pion TFF of ref.~\cite{Leutgeb:2022lqw} (LMR22) with
OPE fit  (blue lines) and  $F_\rho$-fit (red lines) compared to the dispersive results of
ref.~\cite{Hoferichter:2018kwz} (DR)
and experimental data. (Figure adapted from \cite{\WPnew},
includes also the recent data from BESIII \cite{BESIII:2025zjx})}
\label{fig:pi0TFF}       
\end{figure}

\begin{figure}
\centering
\includegraphics[width=0.5\textwidth,clip]{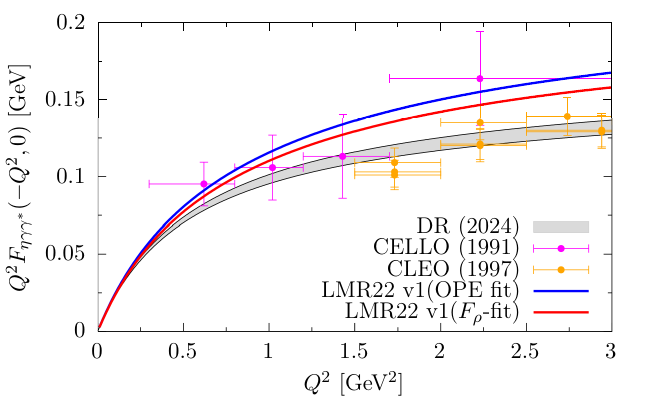}\includegraphics[width=0.5\textwidth,clip]{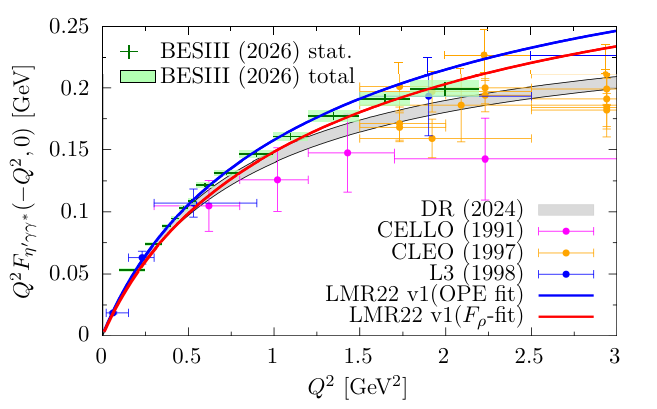}
\caption{Results for the singly virtual
$\eta$ and $\eta'$ TFF of Ref.~\cite{Leutgeb:2022lqw} (LMR22) with
OPE fit  (blue lines) and  $F_\rho$-fit (red lines) compared to the dispersive results of
ref.~\cite{Holz:2024diw} and experimental data (figure adapted from \cite{\WPnew}). Also included in the right panel are the new high-precision BESIII data for $\eta'$ \cite{BESIII:2026bks}
which turn out to be significantly above the dispersive result, very close to and mostly in between the two LMR22 results.}
\label{fig:svetasTFF}       
\end{figure}

Fig.~\ref{fig:pi0TFF} shows the singly and doubly
virtual transition form factors (TFFs) for pions
obtained in the LMR22 model (which for pions is
identical with the HW1m model of ref.~\cite{Leutgeb:2021mpu}), which are in remarkable
agreement with the dispersive results of
ref.~\cite{Hoferichter:2018kwz} as well as the recent data from BESIII \cite{BESIII:2025zjx}.
Fig.~\ref{fig:svetasTFF} displays singly virtual TFFs for
$\eta$ and $\eta'$, where the extension to the Katz-Schwartz
model with nonvanishing gluon condensate in LMR22
is essential to have good agreement with the
two-photon width. The singly virtual results
are in both cases significantly above the most recent dispersive
results of ref.~\cite{Holz:2024diw}. However,
a recent high-precision measurement of the $\eta'$ TFF
in ref.~\cite{BESIII:2026bks} (also included in the right panel of fig.~\ref{fig:svetasTFF})
is in full agreement with the LMR22 results;
for almost all virtualities the experimental values
lie in between the two versions OPE fit and $F_\rho$-fit
that have been used to fix the holographic coupling $g_5$.
(The OPE fit matches the leading-order pQCD limit exactly, while the
$F_\rho$-fit is closer to the OPE result with gluonic corrections
\cite{Bijnens:2021jqo}
at moderately large $Q^2$.)

Another hQCD model, where the U(1)$_A$ anomaly
and finite quark masses have been included, is the
soft-wall model of 
\cite{Colangelo:2023een,Colangelo:2024xfh}.
However, as found in \cite{Leutgeb:2025jmv},
this model cannot be used, at least not in its
original version, to describe the HLbL amplitude
due to divergences.

Holographic QCD models provide a realization of
vector meson dominance (VMD) with an infinite tower
of $\rho,\omega,\phi$ mesons.
Unlike hadronic models with a finite number of these
resonances, one finds an asymptotic behavior
of TFFs in agreement with the
light-cone expansion (LCE) 
results of \cite{Brodsky:1981rp,Hoferichter:2020lap}
for pseudoscalars and axial-vector mesons, to wit, 

\be
                F_{\pi^0\gamma^*\gamma^*}(Q_1^2,Q_2^2)\to\frac{ 2f_\pi}{Q^2}\left[ \frac1{w^2}-\frac{1-w^2}{2w^3}\ln\frac{1+w}{1-w} \right],
                \label{pionTFFas}
\ee 
and \cite{Leutgeb:2019gbz}\footnote{This result was
in fact obtained prior to its derivation in QCD in \cite{Hoferichter:2020lap}.}
\be\label{Aasymptotics}
A_n(Q_1^2,Q_2^2) \to \frac{12\pi^2 F^A_{n}}{N_c Q^4}
                \frac1{w^4}\left[
                w(3-2w)+\frac12 (w+3)(1-w)\ln\frac{1-w}{1+w}
                \right],
\ee
where $w=\frac{Q_1^2-Q_2^2}{Q_1^2+Q_2^2}$ with photon virtualities $Q_i$ and $Q^2=\frac12(Q_1^2+Q_2^2)$.

In \eqref{Aasymptotics}, $n=1,2,\ldots,\infty$ is the mode
number of an
infinite tower of axial-vector mesons, which turn out
to be necessary for saturating the MV constraint \cite{Melnikov:2003xd} on the
longitudinal component of the HLbL amplitude
\begin{equation}
\label{eq:MVConstr}
    \lim_{Q_3\rightarrow \infty}\lim_{Q \rightarrow \infty}  Q_3^2 Q^2 \bar{\Pi}_1(Q,Q,Q_3)= -\frac{2}{3 \pi^2},
\end{equation}
where each individual mode has a vanishing limit and
only the infinite sum is nontrivial
\cite{Leutgeb:2019gbz,Cappiello:2019hwh}.
The infinite tower of pseudoscalar mesons 
does not contribute to \eqref{eq:MVConstr} \cite{Leutgeb:2021mpu};
its contribution is suppressed by a factor $(Q_3/Q)^2\ln(Q_3/Q)$
(see ref.~\cite{Mager:2026fkq} for an analytic proof).

\section{Minimally coupled tensor mesons and completion of SDCs}

Because holographic models are derived from a gravitational theory in
higher dimensions, they naturally include tensor mesons which are dual to metric fluctuations. 
Katz, Lewandowski, and Schwartz \cite{Katz:2005ir} (KLS)
have evaluated their two-photon and two-pion couplings
by matching the energy-momentum correlator to the QCD result
and have found that the resulting two-photon width
as well as the mass of the tensor meson fit
remarkably well to the values of $f_2(1270)$.

In \cite{Mager:2025pvz,Cappiello:2025fyf} we have
worked out the TFFs of the HW tensor mesons and have
found that with the minimal coupling provided by
the diffeomorphism invariant action \eqref{S5}
there are two nonvanishing TFFs, out of five that
are possible for on-shell transverse-traceless tensor mesons.

In contrast to the case of axial-vector mesons, their
asymptotic limits do not agree with the asymmetry functions
obtained in the LC results of \cite{Hoferichter:2020lap},
only their power-law behavior is correct.
(The same is true by construction for the
simple quark-model ansatz of \cite{Schuler:1997yw}.)

\begin{figure}
\centering
\includegraphics[width=0.4\textwidth,clip]{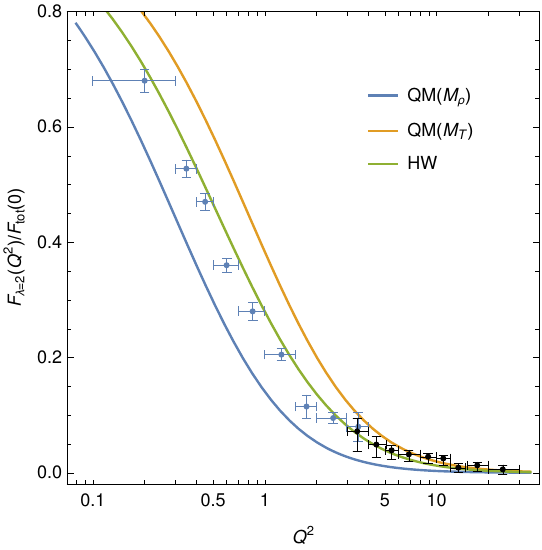}\qquad\includegraphics[width=0.4\textwidth,clip]{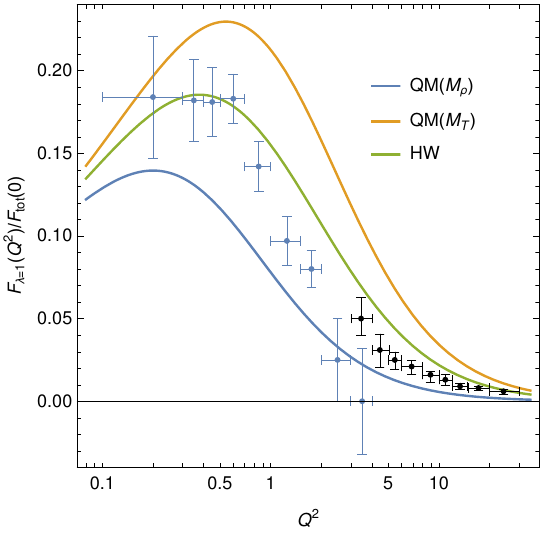}
\includegraphics[width=0.4\textwidth,clip]{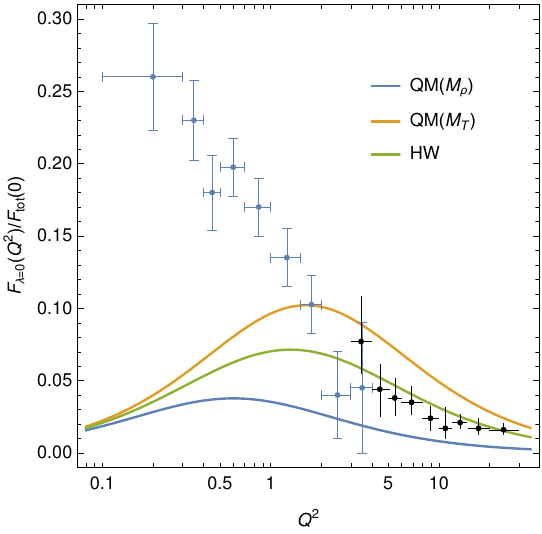}\qquad\includegraphics[width=0.4\textwidth,clip]{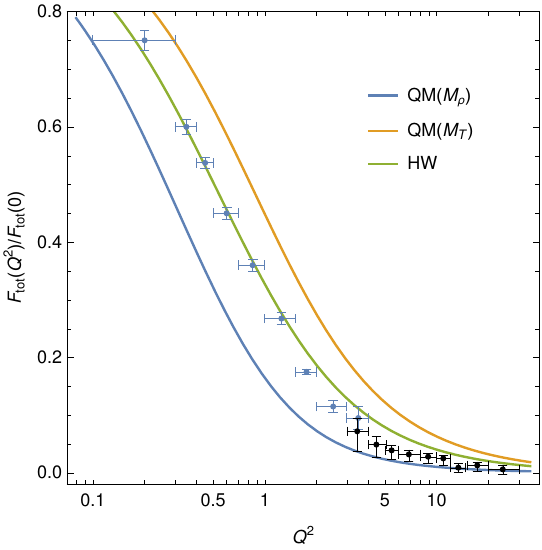}
\caption{Singly virtual tensor helicity amplitudes 
of the $f_2(1270)$ tensor meson
normalized to $F_\mathrm{tot}$
($\lambda=2,1,0$ and tot, their sum) for the HW model and two variants of the simple quark model (QM)
\cite{Schuler:1997yw} 
with scale parameter $\Lambda_T$ set to either $M_\rho$ \cite{Hoferichter:2024bae} or $M_T$ \cite{Hoferichter:2020lap,Schuler:1997yw},
compared to BELLE data \cite{Belle:2015oin} (in black) and preliminary low-$Q^2$
data from BESIII \cite{Lellmann:2026bie} (in blue)}
\label{fig:svtensorTFF}       
\end{figure}

Comparing with experimental data for the singly
virtual helicity amplitudes of the $f_2(1270)$,
one finds, however, that the hQCD result reproduces 
the leading helicity-2 result (first panel in
fig.~\ref{fig:svtensorTFF}) surprisingly well,
much better than the simple quark-model ansatz,
while both are missing the TFF $\F_2^T$ required for
a nonvanishing helicity-0 amplitude for real photons.
Nevertheless, the most recent, but preliminary data
at low $Q^2$ from BESIII \cite{Lellmann:2026bie}
show extremely good agreement for $F_\mathrm{tot}$
(the sum over helicities, displayed in the fourth panel in fig.~\ref{fig:svtensorTFF}).

Considering the full infinite tower of tensor mesons,
we have found that they also contribute to the
short-distance limits of the HLbL amplitude, except
for the longitudinal MV constraint \eqref{eq:MVConstr}, which is saturated exclusively
by the tower of axial-vector mesons.
The symmetric longitudinal SDC is, however,
matched to only 81\% by the latter.
The tower of tensor mesons turns out to have the
correct sign to fill the gap. However, while the
numerical size is approximately right for $N_c=N_f=3$,
it has a wrong $N_c$ scaling.
In \cite{Mager:2025pvz,Cappiello:2025fyf} we have
proposed to replace the choice of $k_T$ of \cite{Katz:2005ir}
such that the symmetric longitudinal SDC is matched
completely. The original choice of $k_T$ is actually
one that corresponds to flavor-singlet tensor glueballs
rather than tensor quarkonia.
The correspondingly rescaled tensor mode is thus
assumed to be the combined effect of a multiplet of the latter.
Distributing the resulting two-photon width of
the lowest mode among
the ground-state mesons $f_2(1270)$, $a_2(1230)$, and $f_2'(1525)$
under the assumption of ideal mixing in fact gives
a near-perfect match with experimental values.

With this set-up, we have evaluated the contribution
to $a_\upmu$ from all the holographic tensor modes,
yielding a relatively large positive contribution of
\begin{align}\label{amufulltower}
    a_\mu^{T}\times 10^{11}=11.1_{-3.0}^{+1.3}
    \qquad[8.5_{-2.3}^{+1.0} \;\text{from}\; Q_i<1.5 \mathrm{GeV}],
\end{align}
while the quark-loop ansatz (for the ground-state tensors)
gives \cite{\HSZ} a negative result of $-2.5(8)$ for $Q_i<1.5$ GeV.

While this holographic result is clearly not
the final answer, since it does not provide
five TFFs with their correct short-distance limits,
it shows that tensor mesons can in principle
give sufficiently sizeable positive contributions such that
the existing tension between lattice and data-driven results
is removed.

\begin{figure}
    \centering
    \includegraphics[width=0.6\linewidth]{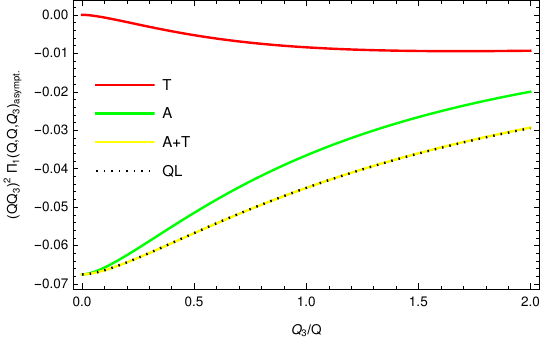}
    \hspace*{2.1mm}\includegraphics[width=0.582\linewidth]{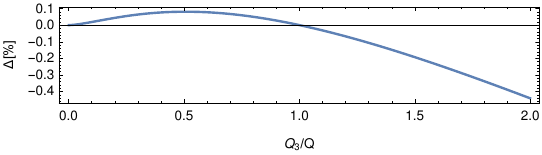}
    \caption{Holographic results for the 
    contribution of the infinite tower of tensor (T) mesons and the one
    of axial-vector (A) mesons to the asymptotic longitudinal function 
    $\hat\Pi_1(Q,Q,Q_3)\times Q^2 Q_3^2$ as a function of $Q_3/Q$, with their sum
    compared to the massless quark-loop (QL) result, the lower panel showing the deviation
    in percent. The MV constraint corresponds to $Q_3/Q\to 0$, which for $Q_i\gg m_\rho$ is saturated exclusively by the axial-vector tower.}
    \label{fig:wow}
\end{figure}

As further evidence that the holographic result may
already be close to one where all short-distance limits
of pQCD are satisfied we show how well the tensor
contributions fill the gap in the longitudinal
structure function $\hat\Pi_1$ in fig.~\ref{fig:wow},
where for asymptotic $Q_1=Q_2\not=Q_3$
the deviation from the leading-order OPE result
in QCD is at the permille level.
In  table \ref{tab:results} we list moreover the
situation in the symmetric short-distance limit
for the remaining transverse 
structure functions of the HLbL amplitude.
As can be seen there, the axial-vector tower result
yields typically much smaller fractions of the OPE
result, which to leading order is given by the
quark loop (QL).
In all cases, the tensor contributions have the
correct sign to reduce the deviation from the QL
results, and with the exception of
$\hat\Pi_{17}$ the magnitudes are such that the sum is
not far from the 100\% achieved in the longitudinal
$\hat\Pi_1$ by construction.
(More details as well as analytic results are
given in the appendix.)

\begin{table}[h]
{\centering
\small
\begin{tabular}{cc|c|cc|cc|r}
\hline
$\hat i$ & $\bar i$ &QL& A & T & A/QL & T/QL & sum[\%] \\
\hline
1 &1,2 & $-\frac{4}{9\pi^2}=-0.04503$ &$C=-0.0365740$ & $-0.0084576$ & 81.22 & 18.78 & 100 \\
4 &3,4 & $\frac49 C=-0.01626$ &$C/6=-0.006096$ & $-0.011442$ & 37.50 & 70.39 & 108 \\
7 &5,6,7 & $\frac29 C=-0.008128$ &$C/6=-0.006096$ & $-0.002564$ & 75.00 & 31.55 & 107 \\
17 &8,9 & $+0.0103245$ &$-C/6=+0.006096$ & $+0.000345$ & 59.04 & 3.34 & 62 \\
39 &10 & $+0.0265796$ &$-C/2=+0.018287$ & $+0.007693$ & 68.80 & 28.94 & 98 \\
\hline
\end{tabular}
}
\caption{
Symmetric $Q\to\infty$ limit results for
$Q^4\bar\Pi_{\le4}$ and $Q^6\bar\Pi_{\ge5}$ with
$\alpha=\alpha_\mathrm{LSDC}$ ($\hat i$ and $\bar i$ refer to the indices
of $\Pih_i$ and $\Pib_i$, respectively; QL to the 
massless quark loop, A and T to the hQCD
results resulting from the infinite towers of axial-vector and tensor mesons)
\label{tab:results}
}
\end{table}

\section{Outlook}

However, because the SDCs are not fulfilled completely, one
should look for additional contributions at leading order in $N_c$
(which thus cannot be provided by the subleading loop contributions
from one-loop Witten diagrams).

There could be such contributions from towers of scalar mesons \cite{Cappiello:2021vzi}, but
they would contribute only to $\hat\Pi_4$ and $\hat\Pi_{17}$ (which
are incidentally the functions with the largest deviations from the
quark-loop). A more promising avenue seems to be higher-derivative
couplings of tensor mesons in the form of nonminimal gravitational
terms involving the flavor-gauge fields, which would contribute
to all structure functions $\hat\Pi_i$.
They would also be able to furnish nontrivial tensor TFFs $\FT_{2,4,5}$
with nontrivial asymptotic limits at the same order
in $N_c$ as $\FT_{1,3}$, to be compared with the asymptotic light-cone
expansion results \cite{Hoferichter:2020lap},
while the new experimental data from BESIII
\cite{Lellmann:2026bie} constrain the tensor TFFs at
low virtualities. This is work in progress.

\section*{Appendix: Analytic short-distance results}

Remarkably, all short-distance limits of the hQCD results
for the HLbL amplitude with $Q_4=0$ and generic $Q_i\gg m_\rho$ as
well as the massless quark-loop, which provides the leading OPE result \cite{Bijnens:2020xnl,Bijnens:2021jqo}, can be expressed compactly in terms of the
triple $K$-integrals defined and discussed in \cite{Bzowski:2013sza,Bzowski:2020kfw}
\be
I_{\alpha\{\beta_1\beta_2\beta_3\}}(Q_1,Q_2,Q_3)=
\int_0^\infty dz\, z^\alpha \prod_{i=1}^3 Q_i^{\beta_i}K_{\beta_i}(Q_i z).
\ee
This includes the scalar one-loop massless triangle Feynman diagram as the
special case 
\be
I_{1\{000\}}=\frac{i}{4\pi^2}\int d^4 k \frac1{(k^2+i\epsilon)((k-q_1)^2+i\epsilon)((k+q_2)^2+i\epsilon)}
\ee
for spacelike momenta $q_i^2=-Q_i^2$. 
This identification follows from the ``magic connection'' \cite{Davydychev:1995mq} with
2-loop massive vacuum sunset diagrams in two dimensions,
which can be recast immediately into triple $K$-integrals \cite{Adams:2013nia,Groote:2018rpb}. On the other hand, the triangle diagram 
can be expressed in terms of the Clausen function 
in the manifestly symmetric form \cite{Lu:1992ny,Bern:1996ka}
\be
I_{1\{000\}}=\int_0^\infty dz\,z\,K_0(Q_1z)K_0(Q_2z)K_0(Q_3z)=\frac{1}{8 A_{123}}\Big[\mathrm{Cl}_2(2\theta_1)+\mathrm{Cl}_2(2\theta_2)+\mathrm{Cl}_2(2\theta_3)\Big],
\ee
where $\theta_1,\theta_2,\theta_3$ are the interior angles of the triangle with sides $Q_1,Q_2,Q_3$ ($\theta_i$ opposite $Q_i$, so $\theta_1+\theta_2+\theta_3=\pi$), and $A_{123}=\frac14\sqrt{-\lambda}$ is the area of that triangle, with $\lambda(Q_1^2,Q_2^2,Q_3^2)$ denoting the K\"all\'en function. Differentiation with respect to the $Q_i$ permits to 
derive analytic results for other triple $K$-integrals.

The Clausen function (by default of order 2) is defined as
\be
\Cl(x)=-\int_0^x dt\ln\left|2\sin\frac{t}{2}\right|
=\sum_{n=1}^\infty \frac{\sin(nx)}{n^2}.
\ee
At the symmetric point $\theta_i=\pi/3$, one has
$\Cl(2\pi/3)=\frac23 V_0$, where
$V_0$ is the maximum value of the Clausen function, $V_0=\Cl(\pi/3)=1.0149416\ldots$.\footnote{$V_0$
is also known as Gieseking's constant \cite{Adams01121998},
because it gives
the volume of the Gieseking manifold, a nonorientable noncompact hyperbolic 3-manifold with finite
volume, which is the smallest 
among all noncompact hyperbolic 3-manifolds. Moreover, $V_0$ is
the maximal volume of a tetrahedron in hyperbolic 3-space with unit curvature radius.}

In terms of triple $K$-integrals,
the contribution of the massless quark loop 
to  $\hat\Pi_1(Q_1,Q_2,Q_3)$,
which has been worked out in \cite{Colangelo:2019uex,Bijnens:2021jqo},
can be written compactly as [for $N_c=N_f=3$ quarks with charge matrix  $\Q=\operatorname{diag}(2/3,-1/3,-1/3)$]
\begin{align}\label{Pi1QL}
\hat\Pi^\mathrm{QL}_1(Q_1,Q_2,Q_3)
=\frac2{3\pi^2}\left[ 
I_{3\{000\}}(Q_1,Q_2,Q_3)-\frac2{Q_3^2}I_{2\{001\}}(Q_1,Q_2,Q_3)
\right].
\end{align}
In the symmetric case $Q_i=Q$, this can be simplified to
\be\label{Pi1QLsym}
\hat\Pi^\mathrm{QL}_1(Q,Q,Q)=-4/({9\pi^2 Q^4}).
\ee
The other nonvanishing 
symmetric 
quark-loop results $\Pih^\mathrm{QL}_{4,7,17,39}(Q,Q,Q)$
all involve the transcendental constant $V_0$ \cite{Colangelo:2019uex,Bijnens:2021jqo}.

The hQCD results depend on additional mass scales,
in particular $z_0^{-1}\propto m_\rho$.
When all $Q_i\gg m_\rho$ (and other mass parameters in the
Lagrangian), the asymptotic component functions can all be expressed in
terms of triple $K$-integrals. For the longitudinal
component, which is numerically the most important one
for the muon anomalous magnetic moment, we have
\begin{align}
\hat\Pi^\mathrm{A}_1(Q_1,Q_2,Q_3) \to& -\frac{1}{\pi^2 Q_3^2}I_{4\{111\}}(Q_1,Q_2,Q_3)\\
\hat\Pi^\mathrm{T}_1(Q_1,Q_2,Q_3) \to& -2\alpha
\left[ I_{4\{010\}}(Q_1,Q_2,Q_3)+I_{4\{100\}}(Q_1,Q_2,Q_3)
\right]
\end{align}
where
$\alpha=\alpha_\mathrm{KLS}=(\mathrm{tr}\, \Q^2)^2/(g_5^4 k_T)=4/(45\pi^2)$
is the choice of Katz, Lewandowski, and Schwartz \cite{Katz:2005ir},
prior to our rescaling to match the symmetric longitudinal
SDC. 


\subsection*{Symmetric short-distance limits}

In the symmetric limit we have
\begin{align}
    Q^4\hat\Pi^\mathrm{A}_1(Q,Q,Q) \to
    & \,C:=-\frac1{\pi^2}I_{4\{111\}}(1,1,1)=
    -\frac2{27\pi^2}\left[ 33 -4\Dl\right]=-0.0365740\ldots,\\
    &\text{with}\quad\Delta^{(1)}=\psi^{(1)}(\tfrac13)-\psi^{(1)}(\tfrac23)=4\sqrt{3}\,\Cl(\pi/3)=7.03172\ldots,
\end{align}
which is about 81\% of the quark-loop result \eqref{Pi1QLsym}.
The constant $C$ appears also in all other
symmetric limits of $\Pi^A_i$,
\begin{align}
    Q^4 \hat\Pi^\mathrm{A}_1 \simeq 
    6 Q^4 \hat\Pi^\mathrm{A}_4 \simeq
    6 Q^6 \hat\Pi^\mathrm{A}_7 \simeq
    -6 Q^6 \hat\Pi^\mathrm{A}_{17} \simeq
    -2 Q^6 \hat\Pi^\mathrm{A}_{39}
    \to C, \quad \text{for}\;Q_i=Q\to\infty.
\end{align}
Curiously, while $\pi^2 Q^4\Pih_1^\mathrm{QL}(Q,Q,Q)=-\frac49$ 
is a rational number, the same constant $C$ appears also in 
the symmetric quark-loop result for $\hat\Pi_{4,7}$,
although with different prefactors:
$Q^4 \hat\Pi^\mathrm{QL}_4(Q,Q,Q) = 2Q^6 \hat\Pi^\mathrm{QL}_7(Q,Q,Q) = 4C/9$.


The asymptotic symmetric tensor contributions can also be 
expressed in terms of $\Dl$, but the $\hat\Pi_i^T$ all involve
different Bessel moments. E.g., for the longitudinal piece, which is the simplest, we have
\be
    Q^4 \hat\Pi_1^T/\alpha \to
    -4I_{4\{100\}}(1,1,1)=-\frac{16}3 I_{3\{000\}}(1,1,1)
    =-\frac{16}3 \frac{\Dl-6}{9}.
\ee
Curiously, the same Bessel moment $I_{3\{000\}}(1,1,1)$
appears also in the symmetric quark-loop result for $\hat\Pi_{17}$.

For the remaining $\hat\Pi_i^T$ we obtain for the symmetric limit
\begin{align}\label{eq:closed}
&Q^4\Pih_4^T\to\tfrac{2\alpha}{243}(16\Dl-213),\quad
Q^6\Pih_7^T\simeq-\tfrac13Q^6\Pih_{39}^T\to-\tfrac{2\alpha}{81}(16\Dl-105), \nonumber\\
&Q^6\Pih_{17}^T\to\tfrac{2\alpha}{243}(64\Dl-447),
\qquad \text{for}\;Q_i=Q\to\infty.
\end{align}

Rescaling $\alpha$ such that the symmetric SDC
of $\Pih_1$ is satisfied exactly leads to
\begin{equation}\label{eq:alsdc}
\alpha_{\rm LSDC}=\frac{4\Dl-27}{8\pi^2(\Dl-6)}=0.013833,\qquad
\frac{\alpha_{\rm LSDC}}{\alpha_\mathrm{KLS}}=\frac{45(4\Dl-27)}{32(\Dl-6)}=1.53596.
\end{equation}

As discussed above, this does not achieve 
exact agreement with the quark-loop result away from
the symmetric point, except for the Melnikov-Vainshtein
limit, which is saturated exclusively by the axial-vector mesons,
with a vanishing tensor contribution. However, the
numerical deviation of $\Pih_1^A+\Pih_1^T$ from the OPE result
for asymptotic $Q_1=Q_2\not=Q_3$ is in the sub-percent
range, as shown in fig.\ \ref{fig:wow}.

\subsection*{Short-distance corner region}

For $Q_1=Q_2=Q\to \infty$ with large but finite $Q_3\gg z_0^{-1}$, 
the dominant contribution to $a_\upmu$ comes from the
Melnikov-Vainshtein SDC, which in hQCD is saturated by
the tower of axial-vector mesons:
\begin{align}
    \hat\Pi_1^\mathrm{QL}(Q,Q,Q_3)\to -\frac{2}{3\pi^2 Q^2 Q_3^2},\quad
    \hat\Pi_1^\mathrm{A}(Q,Q,Q_3)\to 
    \frac{-1}{Q_3^2 Q^2\pi^2} \int_0^\infty dx \, x^3
    [K_1(x)]^2=
    -\frac{2}{3\pi^2 Q^2 Q_3^2}.
\end{align}

Tensor mesons also contribute to $\Pih_1$, but 
in the same limit
their contribution
is suppressed relative to the leading axial term by
a factor $\mathcal{O}\bigl((Q_3^2/Q^2)\ln(Q/Q_3)\bigr)$.
Explicitly, we obtain
\begin{align}\label{Pi1Tmr}
\hat\Pi^T_1(Q,Q,Q_3)/\alpha 
\to & \, 
\frac4{Q^4} \int_0^\infty dx\, x^4 K_0(x) K_1(x)\left( \ln \frac{xQ_3}{2Q} +\gamma_E\right)
= -\frac83\frac{\ln(Q/Q_3)}{Q^4} + \frac{10}{9Q^4}.
\end{align}

In the transverse sector, the quark-loop result has
similarly enhanced contributions proportional to $1/Q_3^2$
in $\Pih_{5,6,10,14,17,39,50,51}$
for $Q_3\ll Q=Q_1=Q_2\to \infty$:
\begin{align}
    & Q^2 \hat\Pi^\mathrm{QL}_{5,6}\simeq Q^4 \hat\Pi^\mathrm{QL}_{10,14}\simeq
    -Q^4 \hat\Pi^\mathrm{QL}_{17,39} \simeq
    -2 Q^4 \hat\Pi^\mathrm{QL}_{50,51} 
    \to -\frac{2}{9\pi^2}\frac{1}{Q_3^2}.
\end{align}
As shown in \cite{Bijnens:2024jgh}, these limits have contributions
from both the axial and the tensor sector. The part
corresponding to the axial sector is indeed exactly
reproduced by
the axial-vector contributions in hQCD, to wit,
\begin{align}
    & Q^2 \hat\Pi^A_{5,6}\simeq Q^4 \hat\Pi^A_{10,14}\simeq
    -Q^4 \hat\Pi^A_{17,39,50,51} 
    \to 
    -\frac{1}{6\pi^2}\frac{1}{Q_3^2}.
\end{align}
which corresponds to the part of the OPE result 
in \cite{Bijnens:2024jgh}
involving $\omega_{L,T}$ with
$\omega_T\simeq \omega_L/2\to -1/Q_3^2$.

The hQCD tensor contributions read
\begin{align}
    Q^2 \hat\Pi^T_{5,6}\simeq Q^4 \hat\Pi^T_{10,14,50,51}\simeq
    -Q^4 \hat\Pi^T_{39} 
    \simeq 3Q^4 \hat\Pi^T_{17}
    \to -\frac{\alpha}{3Q_3^2}.
\end{align}
If these tensor contributions were scaled up by another factor of $\approx 1.22$, 
 such that $\alpha=\frac1{6\pi^2}=\frac{15}{8}\alpha_\mathrm{KLS}$, 
 one could match all $1/Q_3^2$ enhanced corner limits of the
 quark-loop results except for $\Pih_{17}$
 (where the hQCD tensor contribution has not only the wrong size
 but also the wrong sign to be able to restore the quark-loop result).
However, such a rescaling would lead to an
 overshoot of all symmetric SDCs (except, again, for $\Pih_{17}$).
 
 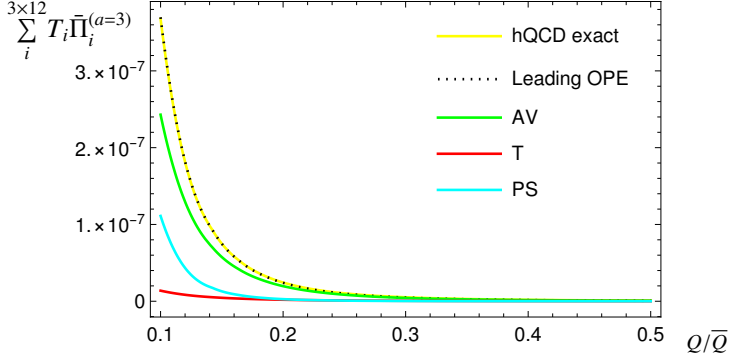
\begin{figure}
    \centering
    \includegraphics[width=0.6\linewidth]{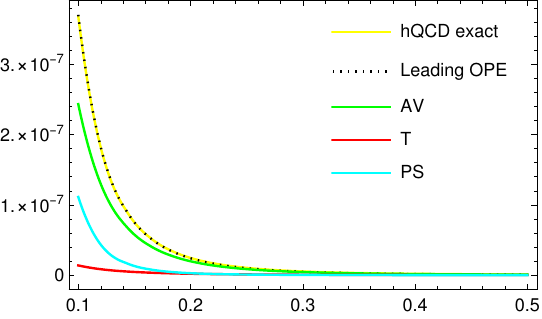}
\begin{picture}(0,0)
\put(5,0){\footnotesize $Q/\overline{Q}$}
\put(-250,117){$\sum\limits_i^{3\times 12}\! T_i\bar\Pi_i^{(a=3)}$}
\end{picture}
    \caption{Adding the three corner contributions for one smaller virtuality $Q$ and two larger ones with $\overline{Q}$, the sum of the latter, fixed at 10 GeV, as plotted in fig.~6 ($y=0$) of \cite{Bijnens:2024jgh}, using the full hQCD result and its contribution from axial vector (AV), pseudoscalar (PS), and tensor mesons (T) compared to the leading OPE result with the holographic results for $\omega_{T,L}$. In agreement with the quark-loop results in \cite{Bijnens:2024jgh}, tensor contributions are strongly suppressed,
    while contributing significantly to the individual $\bar\Pi_i$.}
    \label{fig:BFig6}
\end{figure}
 
 As far as the $a_\upmu$ contribution is concerned,
 however, the symmetric short-distance limits of the tensor contributions
 are presumably more important.
 To leading order in the OPE,
 the tensor contributions in the kinematic corner regions
 (one small virtuality and two large ones)
 all cancel ``miraculously'' \cite{Bijnens:2024jgh}
 when the corner integrands
 are summed and symmetrized (see eq.\ (4.30) in \cite{Bijnens:2024jgh},
 which only involves $\omega_{T,L}$).
 We have verified that this indeed holds
 for the hQCD results. Hence, even though the tensor sector
 typically contributes about 30\% of the axial-sector part of the
 asymptotic limits of individual $\bar\Pi_i$'s, the
 tensor contributions are negligible in asymptotic corner regions,
 as shown numerically in fig.~\ref{fig:BFig6}.

\subsection*{Acknowledgments}
We thank Max Lellmann for helpful correspondence.
This work was funded in part by the Austrian Science Fund (FWF), grant-DOI \url{https://www.doi.org/10.55776/PAT7221623}.  L.C. acknowledges the support of the INFN research project
ENP (Exploring New Physics).

%
\bibliography{SM}

\begin{thebibliography}{47}

\bibitem{Muong-2:2025xyk}
D.P. Aguillard et~al. (Muon g-2), {Measurement of the Positive Muon Anomalous
  Magnetic Moment to 127~ppb}, Phys. Rev. Lett. \textbf{135}, 101802 (2025),
  \texttt{2506.03069}. \doiwoc{10.1103/7clf-sm2v}

\bibitem{Aliberti:2025beg}
R.~Aliberti et~al., {The anomalous magnetic moment of the muon in the Standard
  Model: an update}, Phys. Rept. \textbf{1143}, 1 (2025), \texttt{2505.21476}.
  \doiwoc{10.1016/j.physrep.2025.08.002}

\bibitem{Aoyama:2020ynm}
T.~Aoyama et~al., {The anomalous magnetic moment of the muon in the Standard
  Model}, Phys. Rept. \textbf{887}, 1 (2020), \texttt{2006.04822}.
  \doiwoc{10.1016/j.physrep.2020.07.006}

\bibitem{Boccaletti:2024guq}
A.~Boccaletti et~al., {Hybrid calculation of hadronic vacuum polarization in
  muon $g-2$ to 0.48{\%}}, Nature \textbf{653}, 373 (2026),
  \texttt{2407.10913}. \doiwoc{10.1038/s41586-026-10449-z}

\bibitem{Leutgeb:2022lqw}
J.~Leutgeb, J.~Mager, A.~Rebhan, {Hadronic light-by-light contribution to the
  muon $g-2$ from holographic QCD with solved $U(1)_A$ problem}, Phys. Rev. D
  \textbf{107}, 054021 (2023), \texttt{2211.16562}.
  \doiwoc{10.1103/PhysRevD.107.054021}

\bibitem{Ludtke:2024ase}
J.~L\"udtke, M.~Procura, P.~Stoffer, {Dispersion relations for the hadronic VVA
  correlator}, JHEP \textbf{04}, 130 (2025), \texttt{2410.11946}.
  \doiwoc{10.1007/JHEP04(2025)130}

\bibitem{Hoferichter:2024bae}
M.~Hoferichter, P.~Stoffer, M.~Zillinger, {Dispersion relation for hadronic
  light-by-light scattering: subleading contributions}, JHEP \textbf{02}, 121
  (2025), \texttt{2412.00178}. \doiwoc{10.1007/JHEP02(2025)121}

\bibitem{Schuler:1997yw}
G.A. Schuler, F.A. Berends, R.~van Gulik, {Meson photon transition form-factors
  and resonance cross-sections in $e^+ e^-$ collisions}, Nucl. Phys. B
  \textbf{523}, 423 (1998), \texttt{hep-ph/9710462}.
  \doiwoc{10.1016/S0550-3213(98)00128-X}

\bibitem{Danilkin:2016hnh}
I.~Danilkin, M.~Vanderhaeghen, {Light-by-light scattering sum rules in light of
  new data}, Phys. Rev. D \textbf{95}, 014019 (2017), \texttt{1611.04646}.
  \doiwoc{10.1103/PhysRevD.95.014019}

\bibitem{Cappiello:2025fyf}
L.~Cappiello, J.~Leutgeb, J.~Mager, A.~Rebhan, {Tensor meson transition form
  factors in holographic QCD and the muon g {\ensuremath{-}} 2}, JHEP
  \textbf{07}, 033 (2025), \texttt{2501.09699}.
  \doiwoc{10.1007/JHEP07(2025)033}

\bibitem{Mager:2025pvz}
J.~Mager, L.~Cappiello, J.~Leutgeb, A.~Rebhan, {Longitudinal Short-Distance
  Constraints on Hadronic Light-by-Light Scattering and Tensor-Meson
  Contributions to the Muon $g-2$}, Phys. Rev. Lett. \textbf{135}, 091901
  (2025), \texttt{2501.19293}. \doiwoc{10.1103/dxwr-gpsl}

\bibitem{Estrada:2025bty}
E.J. Estrada, P.~Roig, {Tensor Meson Pole contributions to the HLbL piece of
  $a_{\mu}^{\rm{HLbL}}$ within R$\chi$T}, JHEP \textbf{01}, 070 (2026),
  \texttt{2504.00448}. \doiwoc{10.1007/JHEP01(2026)070}

\bibitem{Melnikov:2003xd}
K.~Melnikov, A.~Vainshtein, {Hadronic light-by-light scattering contribution to
  the muon anomalous magnetic moment revisited}, Phys. Rev. D \textbf{70},
  113006 (2004), \texttt{hep-ph/0312226}. \doiwoc{10.1103/PhysRevD.70.113006}

\bibitem{Colangelo:2019lpu}
G.~Colangelo, F.~Hagelstein, M.~Hoferichter, L.~Laub, P.~Stoffer,
  {Short-distance constraints on hadronic light-by-light scattering in the
  anomalous magnetic moment of the muon}, Phys. Rev. D \textbf{101}, 051501
  (2020), \texttt{1910.11881}. \doiwoc{10.1103/PhysRevD.101.051501}

\bibitem{Colangelo:2019uex}
G.~Colangelo, F.~Hagelstein, M.~Hoferichter, L.~Laub, P.~Stoffer, {Longitudinal
  short-distance constraints for the hadronic light-by-light contribution to
  $(g-2)_\mu$ with large-$N_c$ Regge models}, JHEP \textbf{03}, 101 (2020),
  \texttt{1910.13432}. \doiwoc{10.1007/JHEP03(2020)101}

\bibitem{Bijnens:2020xnl}
J.~Bijnens, N.~Hermansson-Truedsson, L.~Laub, A.~Rodr\'\i{}guez-S\'anchez,
  {Short-distance HLbL contributions to the muon anomalous magnetic moment
  beyond perturbation theory}, JHEP \textbf{10}, 203 (2020),
  \texttt{2008.13487}. \doiwoc{10.1007/JHEP10(2020)203}

\bibitem{Bijnens:2021jqo}
J.~Bijnens, N.~Hermansson-Truedsson, L.~Laub, A.~Rodr\'\i{}guez-S\'anchez, {The
  two-loop perturbative correction to the $(g-2)_\mu$ HLbL at short distances},
  JHEP \textbf{04}, 240 (2021), \texttt{2101.09169}.
  \doiwoc{10.1007/JHEP04(2021)240}

\bibitem{Bijnens:2024jgh}
J.~Bijnens, N.~Hermansson-Truedsson, A.~Rodr\'\i{}guez-S\'anchez, {Constraints
  on the hadronic light-by-light tensor in corner kinematics for the muon g
  \ensuremath{-} 2}, JHEP \textbf{03}, 094 (2025), \texttt{2411.09578}.
  \doiwoc{10.1007/JHEP03(2025)094}

\bibitem{Erlich:2005qh}
J.~Erlich, E.~Katz, D.T. Son, M.A. Stephanov, {QCD and a holographic model of
  hadrons}, Phys. Rev. Lett. \textbf{95}, 261602 (2005),
  \texttt{hep-ph/0501128}. \doiwoc{10.1103/PhysRevLett.95.261602}

\bibitem{Sakai:2004cn}
T.~Sakai, S.~Sugimoto, {Low energy hadron physics in holographic QCD}, Prog.
  Theor. Phys. \textbf{113}, 843 (2005), \texttt{hep-th/0412141}.
  \doiwoc{10.1143/PTP.113.843}

\bibitem{Katz:2007tf}
E.~Katz, M.D. Schwartz, {An Eta primer: Solving the U(1) problem with AdS/QCD},
  JHEP \textbf{08}, 077 (2007), \texttt{0705.0534}.
  \doiwoc{10.1088/1126-6708/2007/08/077}

\bibitem{Leutgeb:2024rfs}
J.~Leutgeb, J.~Mager, A.~Rebhan, {Superconnections in AdS/QCD and the hadronic
  light-by-light contribution to the muon $g-2$}, Phys. Rev. D \textbf{111},
  114001 (2025), \texttt{2411.10432}. \doiwoc{10.1103/PhysRevD.111.114001}

\bibitem{Hoferichter:2018kwz}
M.~Hoferichter, B.L. Hoid, B.~Kubis, S.~Leupold, S.P. Schneider, {Dispersion
  relation for hadronic light-by-light scattering: pion pole}, JHEP
  \textbf{10}, 141 (2018), \texttt{1808.04823}.
  \doiwoc{10.1007/JHEP10(2018)141}

\bibitem{BESIII:2025zjx}
M.~Ablikim et~al. (BESIII), {Measurement of the space-like $\pi^0$ transition
  form factor} (2025), \texttt{2509.07685}.

\bibitem{Leutgeb:2021mpu}
J.~Leutgeb, A.~Rebhan, {Hadronic light-by-light contribution to the muon $g-2$
  from holographic QCD with massive pions}, Phys. Rev. D \textbf{104}, 094017
  (2021), \texttt{2108.12345}. \doiwoc{10.1103/PhysRevD.104.094017}

\bibitem{Holz:2024diw}
S.~Holz, M.~Hoferichter, B.L. Hoid, B.~Kubis, {Dispersion relation for hadronic
  light-by-light scattering: \ensuremath{\eta} and \ensuremath{\eta}$'$ poles},
  JHEP \textbf{04}, 147 (2025), \texttt{2412.16281}.
  \doiwoc{10.1007/JHEP04(2025)147}

\bibitem{BESIII:2026bks}
M.~Ablikim et~al. (BESIII), {High-precision measurement of the space-like
  $\eta'$ transition form factor} (2026), \texttt{2608.12793}.

\bibitem{Colangelo:2023een}
P.~Colangelo, F.~Giannuzzi, S.~Nicotri, {$\pi^0$, $\eta$, $\eta'$ two-photon
  transition form factors in the holographic soft-wall model and contributions
  to $(g-2)_\mu$}, Phys. Lett. B \textbf{840}, 137878 (2023),
  \texttt{2301.06456}. \doiwoc{10.1016/j.physletb.2023.137878}

\bibitem{Colangelo:2024xfh}
P.~Colangelo, F.~Giannuzzi, S.~Nicotri, {Hadronic light-by-light scattering
  contributions to $(g-2)_{\mu}$ from axial-vector and tensor mesons in the
  holographic soft-wall model}, Phys. Rev. D \textbf{109}, 094036 (2024),
  \texttt{2402.07579}. \doiwoc{10.1103/PhysRevD.109.094036}

\bibitem{Leutgeb:2025jmv}
J.~Leutgeb, J.~Mager, A.~Rebhan, {Divergences in the hadronic light-by-light
  amplitude of the holographic soft-wall model}, Phys. Rev. D \textbf{114},
  036033 (2026), \texttt{2511.11797}. \doiwoc{10.1103/4svg-x1ks}

\bibitem{Brodsky:1981rp}
S.J. Brodsky, G.P. Lepage, {Large Angle Two Photon Exclusive Channels in
  Quantum Chromodynamics}, Phys. Rev. D \textbf{24}, 1808 (1981).
  \doiwoc{10.1103/PhysRevD.24.1808}

\bibitem{Hoferichter:2020lap}
M.~Hoferichter, P.~Stoffer, {Asymptotic behavior of meson transition form
  factors}, JHEP \textbf{05}, 159 (2020), \texttt{2004.06127}.
  \doiwoc{10.1007/JHEP05(2020)159}

\bibitem{Leutgeb:2019gbz}
J.~Leutgeb, A.~Rebhan, {Axial vector transition form factors in holographic QCD
  and their contribution to the anomalous magnetic moment of the muon}, Phys.
  Rev. D \textbf{101}, 114015 (2020), \texttt{1912.01596}.
  \doiwoc{10.1103/PhysRevD.101.114015}

\bibitem{Cappiello:2019hwh}
L.~Cappiello, O.~Cat{\`a}, G.~D'Ambrosio, D.~Greynat, A.~Iyer, {Axial-vector
  and pseudoscalar mesons in the hadronic light-by-light contribution to the
  muon $(g-2)$}, Phys. Rev. D \textbf{102}, 016009 (2020), \texttt{1912.02779}.
  \doiwoc{10.1103/PhysRevD.102.016009}

\bibitem{Mager:2026fkq}
J.~Mager, Ph.D. thesis, Vienna, Tech. U. (2026),
  \urlstyle{tt}\url{https://doi.org/10.34726/hss.2026.138460}

\bibitem{Katz:2005ir}
E.~Katz, A.~Lewandowski, M.D. Schwartz, {Tensor mesons in AdS/QCD}, Phys. Rev.
  D \textbf{74}, 086004 (2006), \texttt{hep-ph/0510388}.
  \doiwoc{10.1103/PhysRevD.74.086004}

\bibitem{Belle:2015oin}
M.~Masuda et~al. (Belle), {Study of $\pi^0$ pair production in single-tag
  two-photon collisions}, Phys. Rev. D \textbf{93}, 032003 (2016),
  \texttt{1508.06757}. \doiwoc{10.1103/PhysRevD.93.032003}

\bibitem{Lellmann:2026bie}
M.~Lellmann, Ph.D. thesis, Mainz U. (2026),
  \urlstyle{tt}\url{https://doi.org/10.25358/openscience-15143}

\bibitem{Cappiello:2021vzi}
L.~Cappiello, O.~Cat\`a, G.~D'Ambrosio, {Scalar resonances in the hadronic
  light-by-light contribution to the muon (g-2)}, Phys. Rev. D \textbf{105},
  056020 (2022), \texttt{2110.05962}. \doiwoc{10.1103/PhysRevD.105.056020}

\bibitem{Bzowski:2013sza}
A.~Bzowski, P.~McFadden, K.~Skenderis, {Implications of conformal invariance in
  momentum space}, JHEP \textbf{03}, 111 (2014), \texttt{1304.7760}.
  \doiwoc{10.1007/JHEP03(2014)111}

\bibitem{Bzowski:2020kfw}
A.~Bzowski, P.~McFadden, K.~Skenderis, {Conformal correlators as simplex
  integrals in momentum space}, JHEP \textbf{01}, 192 (2021),
  \texttt{2008.07543}. \doiwoc{10.1007/JHEP01(2021)192}

\bibitem{Davydychev:1995mq}
A.I. Davydychev, J.B. Tausk, {[A magic] Connection between certain massive and
  massless diagrams}, Phys. Rev. D \textbf{53}, 7381 (1996),
  \texttt{hep-ph/9504431}. \doiwoc{10.1103/PhysRevD.53.7381}

\bibitem{Adams:2013nia}
L.~Adams, C.~Bogner, S.~Weinzierl, {The two-loop sunrise graph with arbitrary
  masses}, J. Math. Phys. \textbf{54}, 052303 (2013), \texttt{1302.7004}.
  \doiwoc{10.1063/1.4804996}

\bibitem{Groote:2018rpb}
S.~Groote, J.G. K{\"o}rner, {Coordinate space calculation of two- and
  three-loop sunrise-type diagrams, elliptic functions and truncated Bessel
  integral identities}, Nucl. Phys. B \textbf{938}, 416 (2019),
  \texttt{1804.10570}. \doiwoc{10.1016/j.nuclphysb.2018.11.023}

\bibitem{Lu:1992ny}
H.J. Lu, C.A. Perez, {Massless one-loop scalar three point integral and
  associated Clausen, Glaisher and L-functions}, SLAC-PUB-5809 (1992),
  \urlstyle{tt}\url{https://www.slac.stanford.edu/pubs/slacpubs/5750/slac-pub-5809.pdf}

\bibitem{Bern:1996ka}
Z.~Bern, L.J. Dixon, D.A. Kosower, S.~Weinzierl, {One-loop amplitudes for $e^+
  e^- \to \bar q q \bar Q Q$}, Nucl. Phys. B \textbf{489}, 3 (1997),
  \texttt{hep-ph/9610370}. \doiwoc{10.1016/S0550-3213(96)00703-1}

\bibitem{Adams01121998}
C.C. Adams, The newest inductee in the number hall of fame, Mathematics
  Magazine \textbf{71}, 341 (1998). \doiwoc{10.1080/0025570X.1998.11996674}

\end{thebibliography}

\end{document}